\documentstyle[multicol,psfig,aps,prl,amssymb]{revtex}
\begin{document}

\draft
\preprint{MPA 1080}
\title{Big Bang Nucleosynthesis 
with Matter--Antimatter Domains}

\author{Jan B. Rehm\cite{jr} 
and Karsten Jedamzik\cite{kj}}

\address{
Max-Planck-Institut f\"ur Astrophysik, Karl-Schwarzschild-Str. 1,
85748 Garching, Germany
}

\date{\today}
\maketitle

\begin{abstract}
We investigate Big Bang nucleosynthesis (hereafter, BBN)
in a cosmic environment characterised by a distribution 
of small-scale matter--antimatter domains. Production of antimatter domains in a
baryo-asymmetric universe is predicted in some
electroweak baryogenesis scenarios. 
We find that cosmic antimatter domains of size
exceeding the neutron-diffusion length at temperature $T \sim 1$~MeV
significantly affect the light-element production. 
Annihilation of antimatter preferentially occurs on neutrons
such that antimatter domains may yield a reduction
of the $^4$He abundance relative to a standard BBN scenario.
In the
limiting case, all neutrons will be removed before the onset of
light-element production,
and a universe with net baryon number but without production of light
elements results.
In general, antimatter domains spoil agreement between BBN abundance
yields and observationally inferred primordial abundances limits which allows
us to derive limits on their presence in the early universe. 
However, if only small amounts of antimatter
are present, BBN with low deuterium {\it and} low $^4$He, as seemingly
favored by current observational data, is possible.

\end{abstract}

\begin{multicols}{2}
\narrowtext


Big Bang Nucleosynthesis is one of  the furthest back--reaching
cosmological probes available.
By means of comparing the predicted and observationally inferred light
element abundances the cosmic conditions
as early as a few seconds after the Big Bang may be scrutinized.
Nucleosynthesis has thus been used
to constrain, for example, inhomogeneities in the baryon-to-photon
ratio or abundances and properties of decaying particles during the
BBN era (for reviews on non-standard BBN see \cite{MM:93:Sa:96}).

In this Letter we examine the BBN process in a universe where
matter and antimatter are segregated. 
A segregation of matter and antimatter may affect 
BBN abundance yields even if the average segregation scale is very small
in terms of characteristic astrophysical scales.
Such small scales imply annihilation of antimatter well before the
present epoch.
We are therefore particularly interested in the case where the universe contains 
net baryon number,
as for example, through an excess of matter domains over antimatter domains.
A baryo-asymmetric universe filled with a distribution of small-scale 
matter--antimatter domains may arise during an epoch of 
baryogenesis
at the electroweak scale. It has been shown within the minimal supersymmetric 
standard model, and under the assumption of explicit as well as 
spontaneous CP violation, that during a first-order electroweak phase
transition the baryogenesis process may result in individual 
bubbles containing either
net baryon number, or net anti-baryon number\cite{CPR:94}. 
Recently, it has been argued that
pre-existing stochastic (hyper)magnetic fields in the early universe,
in conjunction with an era of electroweak baryogenesis, may
cause the production of regions containing either matter or antimatter
\cite{GS:97ab}. 
In general, any segregation of matter and antimatter either produced on
scales larger than the neutron diffusion length at weak freeze--out,
which corresponds to scales entering the horizon at cosmic
temperatures below $T \,\approx\, 1$ TeV \cite{FJMO:94},
or produced on superhorizon scales during earlier inflationary epochs 
may be subject to
constraint by BBN. 

Nevertheless, the study of matter--antimatter domains present during the
BBN era has to our knowledge not received any prior attention in this
context. Earlier work on this problem \cite{St:76} has concentrated on
the presence of matter--antimatter domains in baryo-symmetric
universes (i.e. Omn\`{e}s-model). In essence, this work showed that
antimatter domains and 
successful BBN mutually exclude each other. 
Another aspect, namely homogeneous injection 
of antiprotons into the primordial plasma after 
the completion of a conventional BBN era 
($T\,\lesssim\, 10$~keV) has been investigated by several authors
\cite{CKS:82:BFPS:84:ENS:85:BBB:88}. 
The conclusion of such studies is that only a small fraction of about 
$\sim \rm{a\ few\ }10^{-3}$ antiprotons per
proton may be injected after the BBN epoch. 
Larger fractions lead to overproduction of
mainly $^3$He via $\bar{p}^4{\rm He}$
annihilation reactions. 

Annihilation of antimatter before, or during, the
BBN era may be an astonishingly quiescent
process. For local net antibaryon-to-photon ratios of the order, or less, 
than the net cosmic average baryon-to-photon ratio, $\eta$, the entropy
release associated with annihilation is only a minute fraction of the
entropy in the photon-pair plasma. Heating of the plasma is therefore
unimportant. In the absence of turbulent fluid motions 
and convective processes mixing of matter
and antimatter should mainly proceed via (anti-) baryon diffusion. 
Neutrons diffuse most easily through the plasma and their 
diffusion length $d_n$ at the epoch of weak freeze-out ($T_W\approx 1$~MeV)
is approximately $d_n\approx 10^4$ cm. This corresponds to a 
comoving scale of 0.1 cm at $T\approx 100$~GeV; in what follows we often
quote length scales comoving to this temperature \cite{remark1}.

We distinguish three different limiting cases of the influence
of antimatter domains on BBN, according to the 
segregation scale of matter and antimatter, or equivalently,
the approximate matter-antimatter annihilation time:
(a) Annihilations before weak freeze--out: No effect on
the light--element abundances because the $n/p$--ratio, which is important
for the determination of light-element abundances, 
is reset to the equilibrium value by the fast weak interactions. 
(b) Annihilations between weak freeze--out ($T\approx 1$~MeV) and the
onset of significant $^4{\rm He}$ synthesis ($T\approx 0.08$~MeV): The
$n/p$--ratio is 
strongly affected and a highly non--standard BBN scenario results.
For reference, the neutron diffusion length scale at $T\approx 0.08$~MeV
is roughly $d_n\approx 5 \times 10^7$~cm corresponding to
a comoving scale of $\sim 20$ cm at $T\approx 100$~GeV.
(c) Annihilations after $^4$He synthesis ($T \lesssim 80$ keV): Production of elements and antielements in their respective domains occurs. Subsequent annihilation of antimatter proceeds mainly via $p$ and $\bar{p}$ diffusion since $n(\bar{n})$ are incorporated into light elements. Due to the small movability of $p(\bar{p})$ in the plasma, this occurs mainly after the completion of BBN ($T \lesssim 10$ keV). The subsequent reprocessing of light element abundances by annihilations may substantially deviate
We are here most interested in the non-standard
BBN scenario of case (b). 

Annihilation of an antinucleon on a nucleon typically results in the production
of a few pions and photons, 
i.e. $\bar{N} N \mapsto {\rm a\, few\,}\pi^{\pm},\pi^0,\gamma's$. Pions may
subsequently decay via $\pi^{-}\mapsto \mu^{-}+\bar{\nu}_\mu$
$\mapsto e^{-}+\bar{\nu}_e+\bar{\nu}_{\mu}+\nu_\mu$ and 
$\pi^0\mapsto 2\gamma$. It is of interest if these pions or their decay
products may alter the abundance yields, either through their
effect on weak freeze-out or by, for example, the photodisintegration of
nuclei. We assume that the number of annihilations per photon
is not much larger than a few $\times \, 10^{-10}$.  
In this case, annihilation generated $\nu_e$'s have negligible effect 
on the $n/p$-ratio since their
number density is more than nine orders of magnitude smaller than the
thermal $\nu_e$'s, which govern the weak equilibrium. The same holds for 
electrons and positrons produced in $\mu$-decay which are quickly thermalized
by their electromagnetic interactions. 
Annihilation-generated $\gamma$-rays
cascade on the background photons via pair production
and inverse Compton scattering, on a time scale rapid compared to the
time scale for photodisintegration of nuclei \cite{photo}. 
This cascade only terminates when individual photons do not
have enough energy to further pair-produce on background photons.
For temperatures $T \,\gtrsim\, 3$~keV, the energy of $\gamma$-rays
below the threshold for $e^{\pm}$-production
does not suffice for the photodisintegration of nuclei.
Pions may perturb the $n/p$-ratio
via charge exchange reactions, $\pi^+ + n\mapsto \pi^0 + p$ and
$\pi^- +p\mapsto \pi^0 + n$, provided they do not decay before. 
A typical inverse survival time for pions
against charge exchange reactions, $(1/n_{\pi})dn_{\pi}/dt \approx
5 \times 10^5 \,(\eta_b/3 \times 10^{-10}) \,(T/{\rm MeV})^3 \,{\rm
s}^{-1}$ \cite{RS:88}, should be compared to the decay
rate of charged pions $\Gamma_{\pi^\pm}\approx 3.8 \times 10^7 {\rm s}^{-1}$, 
yielding a fraction $f\sim 0.01\, (\eta_b/3 \times 10^{-10})\,(T/{\rm
MeV})^3$ of pions which may induce 
$n\mapsto p$ interconversion. Here $\eta_b$ is the local baryon-to-photon
ratio in the annihilation region and it is understood that $f$ is 
not to exceed unity.
Assuming the production of about one pion per annihilation 
and $\eta_b \approx 3 \times 10^{-10}$ it is straightforward to show
that charge exchange reactions may affect $Y_p$ by no more than
$\Delta Y_p\lesssim 2 f R$. This is small for small $R$ and we will henceforth
neglect the influence of pions on BBN abundance yields.

The main effect of the presence of antimatter on BBN yields 
for annihilation between $1\,{\rm MeV}\,\gtrsim\, T\,\gtrsim\, 0.08$~MeV 
and small antimatter fractions, $R$, is through
a preferential annihilation of antimatter on neutrons. 
Note that this effect had already been noted in the seventies \cite{St:76} in the
context of Omn\`{e}s' baryosymmetric cosmology with the conclusion that such
scenarios may conflict with a successful light-element synthesis.
Neutrons diffuse out into the antimatter regions and
annihilate there, whereas protons are confined to the matter regions. This
results in a $n/p$-ratio reduction. 
Antineutrons which diffuse into the matter regions
and annihilate on neutrons, as well as on protons, have a comparatively
weaker effect on the $n/p$-ratio, 
if one assumes the density of antineutrons to be smaller
than the density of neutrons. 
If annihilation occurs below a temperature of $T \,\lesssim\, 1$~MeV, weak
interactions are nearly frozen out and the 
residual interactions cannot compensate for this
annihilation-generated $n/p$-ratio reduction.  
Since the final $^4{\rm He}$ mass fraction $Y_p$ strongly depends on
the $n/p$-ratio at $T\approx 80$~keV, in particular, $Y_p\approx 2(n/p)/(n/p+1)$,  
less $^4{\rm He}$ than in standard BBN is produced 
in scenarios with antimatter. 
Abundance yields of other light isotopes are affected comparatively weaker.

We have performed numerical simulations of the light-element nucleosynthesis
in a universe with matter and antimatter domains.
We have incorporated antinucleons and their weak interactions 
into the inhomogeneous nucleosynthesis code of \cite{JFM:94}. 
This code solves the BBN network coupled to all relevant  
hydrodynamic processes, such as diffusion of baryons, 
photon diffusive heat transport, neutrino heat transport and
late--time hydrodynamic expansion of high-density regions. 
The nuclear reaction network and the thermodynamics are treated as
in the Kawano-code \cite{WFH:67:Ka:92}. Note that antielements are not included in our code.

Initial conditions for the determination of abundance yields are
chosen to approximate a distribution of small, spherical antimatter domains 
immersed within ordinary matter regions. This is accomplished
by performing simulations with radial symmetry of antimatter bubbles
surrounded by a comparatively larger matter shell. We have found that due
to the large annihilation rates most annihilation takes place at the
boundary between matter and antimatter domains
\cite{remark2}. 
In our model, we have assumed constant and equal densities in the
matter shells and antimatter bubbles, $\eta_b = \eta_{\bar{b}}$. Since
we used the same cosmic  spatial average matter density  
$\langle\eta\rangle= (1-f_V)\eta_{b}-f_V\eta_{\bar{b}}$ 
in all simulations, this choice defines a unique
relation between the  fraction of space
in antimatter bubbles, $f_V$, and the cosmic average ratio of antinucleons to 
nucleons $R= f_V\eta_{\bar{b}}/(1-f_V)\eta_b=f_V/(1-f_V)$. 
It is understood that $R$ is taken well after the
QCD-transition so that there is no thermally produced baryon number,
in particular, the $\eta$'s reflect net (anti-) baryon-to-photon ratios.

\begin{figure}
   \centerline{\psfig{figure=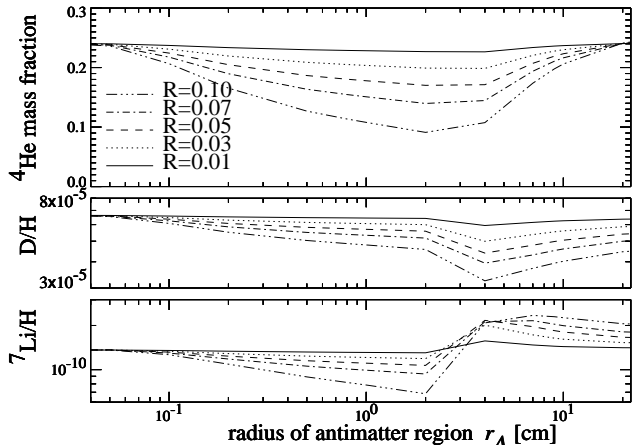,width=3.375in}}
{\caption{Mass fraction of $^4{\rm He}$ and number densities of deuterium and
$^7{\rm Li}$ relative to hydrogen as a function of the radius of the
 antimatter region $r_A$ in comoving units
 \protect\cite{remark1}. Results are shown for different antimatter
 fractions $R$.
  \label{f:all3}}}
\end{figure}

In Figure \ref{f:all3} we show the abundance yields of such matter--antimatter BBN
scenarios as a function of the radius of the antimatter domains, $r_A$, and for
various antimatter fractions, $R$. 
All calculations are performed for average net 
$\langle\eta\rangle = 3.43\times 10^{-10}$, corresponding
to a fractional contribution of baryons to the critical density of
$\Omega_b=0.0125h^{-2}$, with $h$ the Hubble constant in units of 100 km
s$^{-1}$ Mpc$^{-1}$. 

It is evident from the top panel of Figure \ref{f:all3} that 
the presence of antimatter during BBN
results in a reduction of the $^4$He abundance compared to the
abundance in a standard BBN scenario (the values at the far left of
Figure \ref{f:all3}). The effect of $^4$He-reduction is more pronounced for larger
matter/antimatter separation length scales. 
Interconversion of neutrons into protons through residual
weak interactions after the approximate weak freeze-out ($T_W$)
and neutron decay still reduce somewhat the  $n/p$-ratio between
weak freeze-out and synthesis
of $^4$He. The rates for these processes are proportional to the
neutron density. One may approximately assume that a fixed number of 
neutrons are removed through annihilation on antimatter for models with
the same $R$. Early annihilation then weakens the 
post-weak freeze-out reduction of neutron density compared to annihilation
taking place late.
When segregation length scales become too large some annihilation
of antimatter occurs after the binding of neutrons into $^4$He. 
Since this is not yet properly treated in our code, the results on length scales larger than $\approx$ a few cm in Figure \ref{f:all3} may be subject to modification, cf. \cite{HKS}.

Within observational primordial abundance determination uncertainties the
abundances of the other light  
isotopes are only significantly modified with respect to standard BBN for
large antimatter fractions. However, such large antimatter fractions
produce observationally disallowed low $^4$He mass fractions.
It is interesting to note that there is no limit to the amount of $^4$He
reduction possible. In particular, we have found scenarios with large antimatter
fractions where {\it all} neutrons will be removed at temperatures
above $T\approx 80$~keV. In such scenarios there is also 
negligible production of other light elements.
These would result in universes, or local regions of the
universe, where {\it no} light-element production in the early universe occurs,
but which, nevertheless, have net baryon number (cf. \cite{St:76}). 

\begin{figure}
\centerline{\psfig{figure=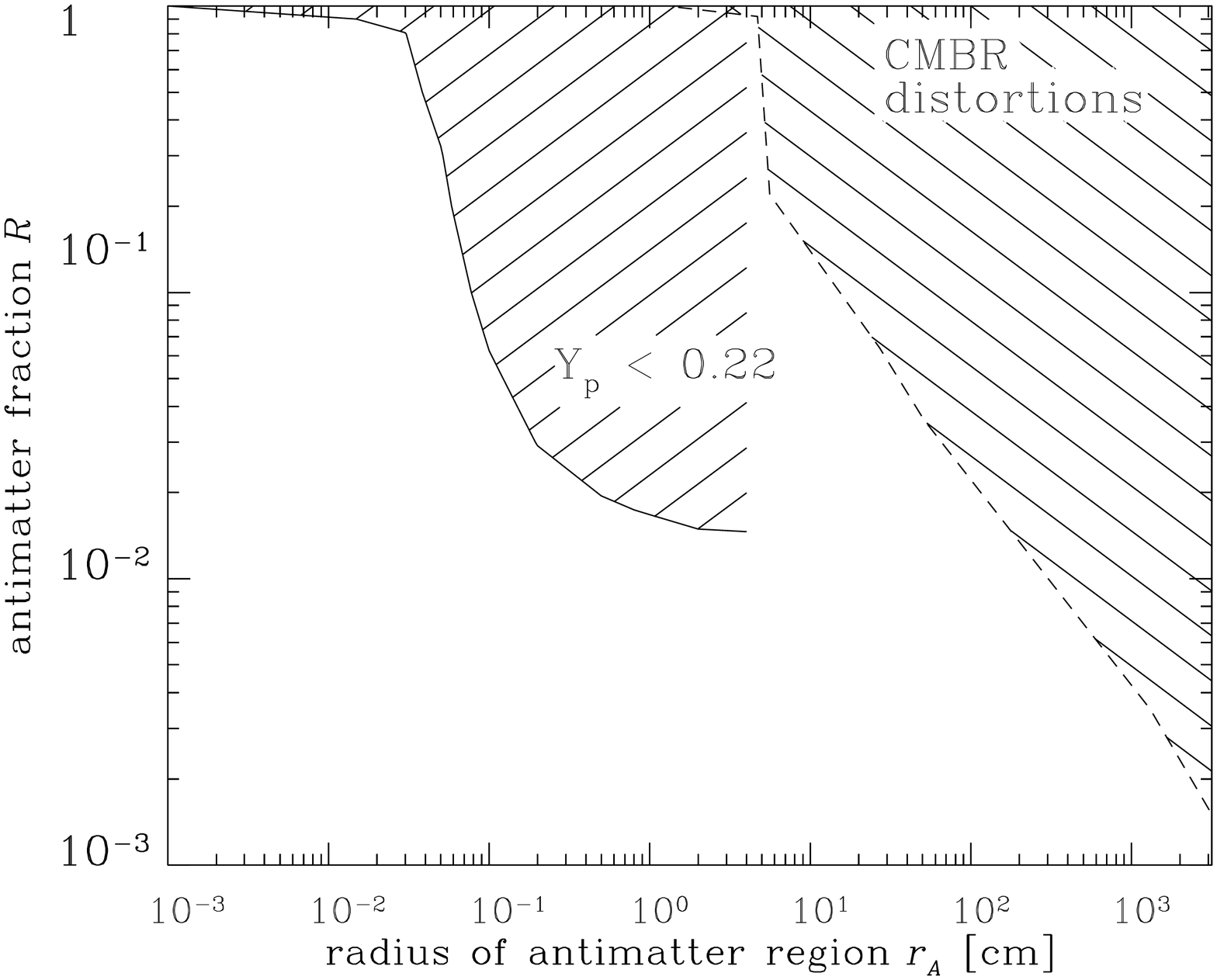,width=3.375in}}
  {\caption{Schematic limits from BBN and CMBR distortions on the ratio
  of antinucleons 
to nucleons, $R$, allowed to be present in the early universe, $r_A$ is again
given in comoving units \protect\cite{remark1}. The constraint for
  large antimatter fractions is somewhat uncertain due to the
  neglected influence of pions, as discussed in the text. \label{f:limits}}}
\end{figure}

In Figure \ref{f:limits} we compare approximate limits on the amount of antimatter from our BBN studies to similar limits derived from distortions of the cosmic microwave background
radiation (CMBR). Whereas for $r_A$ 
between the neutron diffusion length at weak freeze-out and 
onset of $^4$He synthesis the constraints derive from
an underabundant production of $^4$He, for larger $r_A$
limits are due to CMBR distortions.
Input of non--thermal energy into the CMBR will be fully thermalized by
means of Compton and double Compton scattering 
at redshifts larger than $z = 10^7$.
For the computation of the CMBR limits 
a characteristic annihilation redshift $z$ of antimatter domains of size
$r_A$ was computed by the equality, $r_A=d_p(z)$, where $d_p(z)$ is the
proton diffusion length. Limits were then determined by requiring that
the amount of energy released by annihilations into the CMBR 
may not exceed the redshift-dependent
observationally allowed release of nonthermal energy into the CMBR
as given in \cite{FCGMSW:96}. 

The BBN constraints on antimatter may be used, in turn, to constrain
processes occurring in the early universe before the BBN
epoch. In the electroweak baryogenesis scenario by \cite{CPR:94}
individual bubbles 
of broken phase during a first-order electroweak transition create either
net baryon or antibaryon number. Segregation of matter and antimatter is
expected on the electroweak bubbles scale, $l^{EW}_N$, which has
been estimated to fall in the range $l_N^{EW}\sim (10^{-5}-10^{-1})
l_H^{EW}$, with $l_H^{EW}$ the electroweak horizon scale
$l_H^{EW}\approx 2{\rm cm}(T/100 {\rm GeV})^{-2}$ \cite{nucl}. 
Observable
signatures in the BBN yields for such segregation scales are not anticipated.
Nevertheless, some scenarios of \cite{CPR:94} envision only a slight excess in the
number of bubbles, $N_b$, over the number of antibubbles, $\bar{N}_b$,
i.e. $(N_b-\bar{N}_b)/(N_b+\bar{N}_b)\sim 10^{-3}$. 
In this case the effective
segregation scale of a realistic distribution may well exceed the bubble scale.
Meaningful limits on the models by \cite{GS:97ab} from BBN are more probable. 
Here stochastic
matter--antimatter distributions are produced by pre-existing (hyper)magnetic
fields. Segregation scales are not related to $l_N^{EW}$, but to the
coherence scale of the magnetic fields. Assuming equipartition between the
energy in the magnetic fields and radiation, a scale-invariant magnetic field 
distribution, for example, may yield average   
(anti-) baryon-to-photon ratios as large as $|\eta |\sim 10^{-5}$ 
on the scale $d_n(T_W)$. Even though scale-invariance may be
unrealistic, 
a similar scenario with a factor $10^5$ smaller fluctuation
amplitude already leaves an imprint on the BBN abundances.    
In any case, fluctuation amplitudes as large as  $|\eta |\sim 10^{-5}$ 
have not been properly investigated in this Letter due to the
influence of pions.  
It is further difficult to assess and model a realistic three-dimensional
matter--antimatter distribution and its BBN yields.

It is tempting to speculate about BBN yields from stochastic distributions 
with net cosmic $\eta$ larger than that inferred from standard BBN.
Local regions with large $R$ do not contribute to the light
element nucleosynthesis but contribute to $\Omega_b$. 
If the protons of such regions are mixed with the yields of regions at
low $R$ and low as well as high $\eta$, some relaxation of the bound
on $\Omega_b$ inferred from standard  BBN may result. It seems, however,
not easy to satisfy simultaneously observational constraints on all the 
light-element abundances at much larger $\Omega_b$. It is also not
clear if such scenarios would require extreme fine-tuning of fluctuation
parameters.

In summary, we have investigated BBN in the presence of matter--antimatter
domains. We have concentrated on the study of BBN with small
antimatter fractions at the canonical cosmic 
net baryon-to-photon ratio $\sim 3\times
10^{-10}$. Surprisingly, antimatter fractions of the order $R\,\lesssim\, 1\%$
surviving to the epoch of weak freeze-out may considerably alleviate the
long-standing tension between the primordial $^4$He and deuterium  abundances
in standard BBN and their observationally inferred limits
\cite{crisis}. Even though 
other modifications of BBN have been proposed which also yield low
$^4$He and deuterium \cite{criss}, 
our model may have some motivation
from electroweak
baryogenesis scenarios which may generate antimatter domains.

We acknowledge useful discussions with A.~Kercek and G.~Raffelt and want to thank H.~Kurki-Suonio and E.~Sihvola for communcating their results to us prior to publication. We
also appreciate the constructive criticism by G.~Steigman, 
as well as an anonymous referee.
This work was supported in part by the "Sonderforschungsbereich 375-95
f\"ur Astro-Teilchenphysik" der Deutschen Forschungsgemeinschaft.

\end{multicols}

\begin{thebibliography}{10}
\bibitem[*]{jr}
jan@mpa-garching.mpg.de
\bibitem[\dagger]{kj}
jedamzik@mpa-garching.mpg.de

\bibitem{MM:93:Sa:96}
R.A.~Malaney and G.J.~Mathews, Phys. Pep. {\bf 229},  145  (1993);
S.~Sarkar, Rept. Prog. Phys. {\bf 59},  1493  (1996).

\bibitem{CPR:94}
D.~Comelli, M.~Pietroni, and A.~Riotto, Nucl. Phys.~B {\bf 412},  441  (1994).

\bibitem{GS:97ab}
M.~Giovannini and M.E.~Shaposhnikov, Phys. Rev. Lett. {\bf 80}, 22 (1998);
M.~Giovannini and M.E.~Shaposhnikov, Phys. Rev. D {\bf 57}, 2186 (1998).

\bibitem{FJMO:94}
G.M.~Fuller, K.~ Jedamzik, G.J.~Mathews, and A.~Olinto,
     Phys. Lett.~B {\bf 333}, 135 (1994).

\bibitem{St:76}
G.~Steigman, Ann. Rev. Astron. Astrophys. {\bf 14}, 339 (1976).

\bibitem{CKS:82:BFPS:84:ENS:85:BBB:88}
F.~Combes, O.~Fassi-Fehri and B.~Leroy, Nature, {\bf 253}, 25 (1975); 
V.M.~Chechetkin, M.Yu.~Khlopov, and M.G.~Sapozhnikov, Riv.
  Nuovo Cim. {\bf 5},  1  (1982);
Yu.A.~Batusov {\it et~al.}, Nuovo Cim. Lett. {\bf 41},  223
  (1984);
J.~Ellis, D.~V.~Nanopoulos, and S.~Sarkar, Nuc. Phys. B, {\bf 259} 175
(1985); 
R.~Dom\'{\i}nguez-Tenreiro and G.~Yepes, ApJ, {\bf 317}, L1 (1987); 
G.~Yepes and R.~Dom\'{\i}nguez-Tenreiro, ApJ, {\bf 335}, 3 (1988); 
F.~Balestra {\it et~al.}, Nuovo Cim. {\bf 100A},  323  (1988).

\bibitem{remark1}
Proper length $l_p(T)$ at temperature $T$ 
may be obtained from these comoving scales, $l_c$, via $l_p(T)=l_c
\left(g(T)/g(T=100{\rm GeV})\right)^{-1/3}(T/100 
{\rm GeV})^{-1}$, where $g$ denotes the statistical weight of relativistic particles.  

\bibitem{HKS}H.~Kurki-Suonio and E. Sihvola, in preparation.

\bibitem{rj}J.B.~Rehm and K.~Jedamzik, in preparation.


\bibitem{photo}D.~Lindley, MNRAS, {\bf 193}, 593 (1980); J.~Ellis, 
G.B.~Gelmini, J.P.~Lopez, D.V.~Nanopoulos, and S.~Sarkar, Nucl. Phys.~B
{\bf 373}, 399 (1992); R.J.~Protheroe, T.Stanev, and V.S.~Berezinsky,
Phys. Rev.~D {\bf 51}, 4134 (1995).

\bibitem{RS:88}
M.H.~Reno and D.~Seckel, Phys. Rev.~D {\bf 37},  3441  (1988).

\bibitem{JFM:94}
K.~Jedamzik, G.M.~Fuller, and G.J.~Mathews, ApJ {\bf 423},  50  (1994).

\bibitem{WFH:67:Ka:92}
R.V.~Wagoner, W.A.~Fowler, and F.~Hoyle, ApJ {\bf 148},  3  (1967);
L.~Kawano, FERMILAB-Pub-92/04-A  (1992).

\bibitem{remark2}
For the thermally averaged cross sections $\langle\sigma v\rangle$ of the
reactions $n\bar{n},\,p\bar{p},\,n\bar{p},$ and $p\bar{n}$ we used a
value of 40 mb, as suggested by laboratory measurements \cite{anni}. 
The exact value of the cross sections for the $\bar{p}$ reactions are
not relevant for this work, but there may be some dependence of our
results on the value of the $n\bar{n}$ and $p\bar{n}$ cross sections. 
Nevertheless, modified cross sections will not result in significantly
changed abundance yield compared to those shown in this paper.

\bibitem{FCGMSW:96}
D.~J.~Fixsen {\it et~al.}, ApJ {\bf473}, 576 (1996).

\bibitem{anni}
C.~Dover {\it et~al.}, Prog. in Part. and Nucl. Phys. {\bf 29},  87
(1992); 
A.~Bertin {\it et~al.}, Phys. Lett.~B {\bf 369},  77  (1996), (OBELIX
  Collaboration).

\bibitem{nucl}
K.~Enquist, J.~Ignatius, K.~Kajantie, and K.~Rummukainen, Phys. Rev.~D
{\bf 45}, 3415 (1992);
M.~Dine, R.G.~Leigh, P.~Huet, A.~Linde, and D.~Linde, Phys. Rev.~D
{\bf 46}, 550 (1992);
B.H.~Liu, L.~McLerran, and N.~Turok, Phys Rev.~D {\bf 46}, 2668 (1992).

\bibitem{crisis}
C.J.~Copi,  D.N.~Schramm, and M.S.~Turner, Phys. Rev. Lett. {\bf 75}, 3981
(1995); 
N.~Hata et~al., Phys. Rev. Lett. {\bf 75},  (1995) 3977.


\bibitem{criss}
E.~Holtmann, M.~Kawasaki, and T.~Moroi, Phys. Rev. Lett. {\bf 77},  3712
  (1996);
K.~Kohri, M.~Kawasaki, and K.~Sato, astro-ph/9705148  (1997);
K.~Kohri, M.~Kawasaki, and K.~Sato, ApJ {\bf 490},  72  (1997);
S.~Hannestad, Phys. Rev. D {\bf 57}, 2213 (1998).


\end{thebibliography}
\end{document}